\documentclass[twocolumn]{aastex701}
\usepackage{graphicx}
\usepackage{subcaption}
\usepackage{CJK}

\begin{document}
\begin{CJK*}{UTF8}{gbsn}
\title{The Astrometric Resoeccentric Degeneracy: \\ Eccentric Single Planets Mimic 2:1 Resonant Planet Pairs in Astrometry}
\shorttitle{The Astrometric Resoeccentric Degeneracy}
\shortauthors{Yahalomi et al.}

\correspondingauthor{Daniel A. Yahalomi}
\email{yahalomi@mit.edu}

\author[0000-0003-4755-584X]{Daniel A. Yahalomi}
\altaffiliation{Juan Carlos Torres Postdoctoral Fellow}
\affiliation{Center for Computational Astrophysics, Flatiron Institute, 162 Fifth Ave, New York, NY 10010, USA}
\affiliation{Kavli Institute for Astrophysics and Space Research, Massachusetts Institute of Technology, Cambridge, MA 02139, USA}
\email{dyahalomi@flatironinstitute.org}

\author[0000-0003-0834-8645]{Tiger Lu (陆均)}
\altaffiliation{Flatiron Research Fellow}
\affiliation{Center for Computational Astrophysics, Flatiron Institute, 162 Fifth Ave, New York, NY 10010, USA}
\email{tlu@flatironinstitute.org}

\author[0000-0001-5032-1396]{Philip J. Armitage}
\affiliation{Center for Computational Astrophysics, Flatiron Institute, 162 Fifth Ave, New York, NY 10010, USA}
\affiliation{Department of Physics and Astronomy, Stony Brook University, Stony Brook, NY 11794, USA}
\email{parmitage@flatironinstitute.org}

\author[0000-0001-9907-7742]{Megan Bedell}
\affiliation{Center for Computational Astrophysics, Flatiron Institute, 162 Fifth Ave, New York, NY 10010, USA}
\email{mbedell@flatironinstitute.org}

\author[0000-0003-0174-0564]{Andrew R. Casey}
\affiliation{Center for Computational Astrophysics, Flatiron Institute, 162 Fifth Ave, New York, NY 10010, USA}
\affiliation{School of Physics \& Astronomy, Monash University, Clayton 3800, Victoria, Australia}
\email{acasey@flatironinstitute.org}

\author[0000-0003-0872-7098]{Adrian M. Price-Whelan}
\affiliation{Center for Computational Astrophysics, Flatiron Institute, 162 Fifth Ave, New York, NY 10010, USA}
\email{aprice-whelan@flatironinstitute.org}

\author[0000-0002-7670-670X]{Malena Rice}
\affiliation{Department of Astronomy, Yale University, New Haven, CT 06511, USA}
\email{malena.rice@yale.edu}

\begin{abstract}
Detections of long-period giant exoplanets will expand dramatically with \textit{Gaia} Data Release 4 (DR4), but interpreting these signals will require care. We derive the astrometric resoeccentric degeneracy: an astrometric analogue of the well-known radial velocity degeneracy in which a single eccentric planet can mimic two circular planets near a 2:1 period ratio. To first order in eccentricity, the sky-projected motion of a single eccentric orbit decomposes into a fundamental mode and first harmonic with an amplitude proportional to that eccentricity. A pair of coplanar, circular planets in a 2:1 orbital resonance produces the same harmonic structure: the outer planet sets the fundamental mode, while the inner planet supplies an apparent first harmonic. We present a mapping between the harmonic amplitudes and effective eccentricity ($e_\mathrm{eff}$) of a single planet that mimics a 2:1 configuration, demonstrating that $e_\mathrm{eff} = \, 2^{1/3}(M_{p,2}/M_{p,1})$, the masses of the inner and outer planets, respectively. Using simulated \textit{Gaia} data we show that (1) coplanar 2:1 systems are statistically indistinguishable from a single eccentric planet and (2) mutual inclination can break this degeneracy. This bias favors detecting mutually inclined systems, often fingerprints of a dynamically hot history -- traces for processes such as planet-planet scattering or secular chaos. Determining the planetary architectures in which this degeneracy holds will be essential for measuring cool-giant occurrence rates with \textit{Gaia} and for inferring their dynamical evolution histories.

\end{abstract}



\section{Introduction}
\label{intro}

Astrometric detection of exoplanets is of particular interest at the moment as we approach the era of \textit{Gaia} epoch astrometry \citep{Gaia2016}. \textit{Gaia} Data Release 4 (DR4) is scheduled to release the first $\sim$5.5 years of time series astrometry for up to $\sim$2.7 billion sources across the complete sky in December 2026.\footnote{According to \url{https://www.cosmos.esa.int/web/gaia/data-release-4} accessed on November 30, 2025.} \textit{Gaia}'s ability to search for giant exoplanets on distant orbits via the reflex motion they induce on their host stars has long been appreciated \citep{Bernstein1995, Casertano1996, Lattanzi2000, Casertano2008, Perryman2014, Sozzetti2014, Ranalli2018}. Recently, \citet{lammers2025exoplanet} re-investigated the exoplanet yield from \textit{Gaia} DR4 and the subsequent mission-length $\sim$10.5 years of time series astrometry in \textit{Gaia} DR5. This analysis included improvements in our understanding of distant giant-planet occurrence, the local stellar population, and \textit{Gaia} astrometric precision. Via injection recovery tests, \citet{lammers2025exoplanet} demonstrated that \textit{Gaia} will be most sensitive to super-Jupiters ($3-13\,M_{\rm Jup}$) on $2-5\,\mathrm{AU}$ orbits around GKM-type stars ($0.4-1.3\,M_{\odot}$) within $500\,\mathrm{pc}$, predicting $\sim$1,900 planets in DR4 and $\sim$38,000 planets in DR5 with masses and orbital periods determined
to better than 20\%. However, they assumed an underlying population solely composed of single planet systems, which, of course, will not be the true underlying population. An important scientific result from \textit{Gaia} will be precise constraints on cool gas giant occurrence rates and so it is essential to understand degeneracies and multi-modalities that will be present in the \textit{Gaia} DR4 and DR5 datasets in order to not bias these occurrence rate determinations.

Radial-velocity (RV) surveys are known to suffer from a degeneracy in which a system of two planets on nearly circular orbits in or near a 2:1 mean-motion resonance can be well fit by a single planet model with a moderately eccentric orbit -- henceforth called the \textit{resoeccentric degeneracy}. This effect was first analyzed in detail by \citet{AngladaEscude2010}, where it was shown that $\sim$35\% of published single planet eccentric solutions are statistically compatible with circular two-planet 2:1 configurations. A subsequent Bayesian model comparison study on a more selective sample of RV eccentric single systems found similar results -- showing that 25\% of the 60 selected systems are likely to be mischaracterized and are in fact two planet systems at a 95\% confidence level \citep{Boisvert2018}. In the RV case, the Keplerian signal of a single eccentric orbit can be expanded as a sum of a fundamental mode at frequency $n$ and a first harmonic at frequency $2n$ -- with the harmonic amplitude $\simeq e$ times the fundamental.  A two-planet circular 2:1 configuration naturally reproduces the same harmonic structure, as the outer planet mimics the fundamental frequency and the inner planet mimics the first harmonic. 

Here, we derive the astrometric analogue of this effect. We show that, to first order in eccentricity, a single eccentric astrometric orbit also decomposes into a fundamental mode at $n$ and a first harmonic at $2n$ in each sky coordinate. A coplanar, circular 2:1 system produces the same harmonic content, so its astrometric signal lies in the same functional space as that of a single eccentric orbit. Mutual inclination introduces additional degrees of freedom that are not representable by a single orbit and therefore will break the degeneracy in sufficiently precise astrometric data.

The connection between this degeneracy and astrometric observations was predicted in \citet{AngladaEscude2010} --  where it was noted that astrometry could provide ways to uncover eccentric imposters for systems in non-coplanar orientations where the inherent degeneracies will be broken. However, the mathematical derivation was not explicitly presented for the astrometric case; rather, \citet{AngladaEscude2010} points the reader to analyses of frequency decomposition for astrometric detection in \citet{Konacki2002}. 

We have organized this paper as follows. In \S\ref{sec:math} we derive the astrometric degeneracy between circular coplanar 2:1 resonant systems and single eccentric orbits via fundamental and first harmonic frequency analysis (good to first order in eccentricity). In \S\ref{sec:simulations} we demonstrate this degeneracy via simulated \textit{Gaia} DR4 ($\sim$5.5 years) and DR5 ($\sim$10.5 years) data. In \S\ref{sec:implications} we discuss the implications of this degeneracy in astrometric datasets, such as \textit{Gaia}, and its impact on interpreting the demographics that arise from \textit{Gaia} as well as the dynamical implications of potential detections of mutually inclined 2:1 resonant planets.

\section{The Resoeccentric Degeneracy}
\label{sec:math}

Here we derive the astrometric resoeccentric degeneracy. In this derivation, we adopt the orbital coordinate frame as described in \citet{Feng2024} and assume purely Keplerian orbits.

The apparent astrometric displacement can be described in terms of the scaled Thiele-Innes constants $A',B',F',G'$ \citep{Thiele1883}, defined as

\begin{eqnarray}
A' &=&\cos\Omega \cos\omega - \sin\Omega \sin\omega \cos i, \\
B' &=& \sin\Omega \cos\omega + \cos\Omega \sin\omega \cos i, \\
F' &=& -\cos\Omega \sin\omega - \sin\Omega \cos\omega \cos i, \\
G' &=& -\sin\Omega \sin\omega + \cos\Omega \cos\omega \cos i,
\end{eqnarray}
where $\Omega, \omega, i$ are the familiar orbital elements: $\Omega$ the longitude of ascending node, $\omega$ the argument of periastron of the star, and $i$ the inclination of the planet.

The stellar reflex motion projected onto the sky plane of a star induced by a single companion planet can be written, in RA $(\Delta\alpha^\ast = \Delta\alpha\cos\delta)$ and declination $(\Delta\delta)$ space, as

\begin{eqnarray}
\Delta\alpha^\ast(t) = \frac{Y(t)}{d} = \frac{B'\,x(t) + G'\,y(t)}{d}, \label{eq:alpha}\\
\Delta\delta(t) = \frac{X(t)}{d} = \frac{A'\,x(t) + F'\,y(t)}{d}, \label{eq:delta}
\end{eqnarray}
where $d$ is the distance to the star.  Here, $x(t)$ and $y(t)$ are coordinates in the orbital plane, before any projection onto the sky. $X(t)$ and $Y(t)$ are the projected sky-plane coordinates of the star's reflex motion (in the local tangent-plane frame, before scaling by distance).

\subsection{Single moderately eccentric orbit: harmonic structure}
\label{sec:single_ecc}

In the orbital plane, the coordinates of the stellar reflex motion due to a single planet are
\begin{eqnarray}
x(t)
&=&
a_\star (\cos E(t) - e) \nonumber, \label{eq:X} \\
x(t)&\approx&
{}a_\star\left[\cos M - \frac{3}{2}e + \frac{1}{2}e\cos 2M\right],
\\[6pt]
y(t)
&=&
a_\star \sqrt{1-e^2}\,\sin E(t), \nonumber \label{eq:Y} \\
y(t)&\approx&
{}a_\star\left[\sin M + \frac{1}{2}e\sin 2M\right],
\end{eqnarray}
where we have solved Kepler's equation to first order in $e$. Here, $a_\star$ is the stellar reflex semi-major axis ($a_\star = \frac{M_p}{M_p+M_\star}\,a$), with $M_p$ and $M_\star$ the planet and stellar masses, respectively, and $a$ the semi-major axis of the planet with respect to the star. $E(t)$ is the eccentric anomaly, $e$ is the eccentricity, $M \equiv nt$ is the mean anomaly and $n = 2\pi/P$ is the mean motion. Plugging into \autoref{eq:alpha} and \autoref{eq:delta},

\begin{eqnarray}
\Delta\alpha^\ast(t)
&\approx&
\frac{a_\star}{d}\!\left[B'\cos M + G'\sin M\right]
\nonumber \\
&&
{}+\, \frac{1}{2}e\,\frac{a_\star}{d}
\!\left[B'\cos 2M + G'\sin 2M\right]
\nonumber \\
&&
{}-\, \frac{3}{2}e\,\frac{a_\star}{d} B',
\\[6pt]
\Delta\delta(t)
&\approx&
\frac{a_\star}{d}\!\left[A'\cos M + F'\sin M\right]
\nonumber \\
&&
{}+\, \frac{1}{2}e\,\frac{a_\star}{d}
\!\left[A'\cos 2M + F'\sin 2M\right]
\nonumber \\
&&
{}-\, \frac{3}{2}e\,\frac{a_\star}{d} A'.
\end{eqnarray}
Each coordinate of reflex motion thus consists of:

\begin{enumerate}
\item a fundamental mode at frequency $n$,
\item a first harmonic at $2n$,
\item a constant (time-invariant) term,
\end{enumerate}
with the harmonic amplitudes proportional to $e$.  The constant terms represent a shift of the apparent orbit center and are degenerate with the fitted mean position; the time-dependent part of the signal is therefore fully described, to first order, by the fundamental mode and first harmonic. For our purposes, these constant terms can be safely discarded.

\subsection{Two circular planets in a 2:1 coplanar configuration}

Now consider two planets on circular orbits around the same star with periods $P_1 = 2P_2$ and stellar reflex semi-major axes $a_{\star,1}$ and $a_{\star,2}$.  We assume that the orbits are coplanar and share the same $(\Omega,i,\omega)$, so that the same scaled Thiele-Innes constants $(A',B',F',G')$ apply to both planets. Assuming the outer planet shares an orbital period with the single planet analyzed in \S \ref{sec:single_ecc} so $P_1 = P$, in the orbital plane the total displacement may be written as a linear combination of contributions from each planet to zeroth order in eccentricity, as we assume circular orbits, or $e_1=e_2=0$:
\begin{eqnarray}
x_{\rm tot}(t) &
\approx a_{\star,1}\cos M + a_{\star,2}\cos 2M, \\
y_{\rm tot}(t) &
\approx a_{\star,1}\sin M + a_{\star,2}\sin 2M.
\end{eqnarray}
Applying the Thiele-Innes relations to the combined motion, we obtain

\begin{eqnarray}
\Delta\alpha^\ast_{\rm tot}(t)
&\approx&
\frac{a_{\star,1}}{d}\!\left[B'\cos M + G'\sin M\right]
\nonumber \\
&&
{}+\, \frac{a_{\star,2}}{d}\!\left[B'\cos 2M + G'\sin 2M\right],
\\[6pt]
\Delta\delta_{\rm tot}(t)
&\approx&
\frac{a_{\star,1}}{d}\!\left[A'\cos M + F'\sin M\right]
\nonumber \\
&&
{}+\, \frac{a_{\star,2}}{d}\!\left[A'\cos 2M + F'\sin 2M\right].
\end{eqnarray}
Thus, just as in the single-eccentric case, each coordinate consists of:

\begin{enumerate}
\item a fundamental mode at frequency $n$,
\item a first harmonic at $2n$.
\end{enumerate}
There is no time-invariant offset here because we have not included an explicit mean position term; in practice such a term is always fitted separately and does not affect the comparison of the time-dependent harmonics.

We note that systems in exact mean motion resonant configurations will be influenced by \textit{N}-body interactions, causing additional effects that may differentiate the signals from Keplerian orbits \citep{beague2013resonance, batygin2013analytic}. For systems in \textit{near} 2:1 mean motion resonance -- a common detected configuration in exoplanet architectures \citep[e.g.,][]{Fabrycky2014} -- this Keplerian approximation should hold \citep{MurrayDermott1999}.

\subsection{Effective eccentricity mapping}

We now make the equivalence between the two descriptions explicitly by comparing the amplitudes of the fundamental and first harmonic terms. This yields the relations

\begin{eqnarray}
a_{\star,1} &=& a_\star \\
a_{\star,2} &=& \frac{1}{2} e_{\rm eff}\, a_\star,
\end{eqnarray}
where $e_{\rm eff}$ is the effective eccentricity of the single planet model. Solving for $e_{\rm eff}$,
\begin{equation}
\label{eq:eeff}
e_{\rm eff} = 2\,\frac{a_{\star,2}}{a_{\star,1}}.
\end{equation}
In other words, the ratio of the astrometric semi-major axes induced by the inner and outer planets sets the effective eccentricity of the single planet orbit that reproduces the same time-dependent harmonic content (up to first order in $e$ and ignoring the constant offset).  Because the Thiele-Innes constants for the two planets are identical in the coplanar, aligned case, this mapping holds simultaneously in both coordinates.

Thus, a coplanar, phase-aligned circular 2:1 configuration produces, to first order, the same astrometric harmonic content as a single eccentric orbit at the outer period, with an effective eccentricity given by \autoref{eq:eeff}.

\subsection{From harmonic amplitudes to effective eccentricity}
The astrometric resoeccentric degeneracy can be related to physical quantities of interest by converting the effective eccentricity $e_{\rm eff}$ into a relation involving the planet masses and periods. The astrometric semi-major axis induced by planet $k$ for a circular orbit is \citep{Sozzetti2005}
\begin{equation}
a_{\star,k} = \frac{M_{p,k}}{M_\star}\,\frac{a_k}{d},
\end{equation}
where $M_{p,k}$ is the planet mass, $a_k$ is the orbital semi-major axis, $M_\star$ is the stellar mass, and $d$ is the distance from the observer to the system.  Using Kepler's third law, the ratio of the two astrometric amplitudes is
\begin{equation}
\frac{a_{\star,2}}{a_{\star,1}}
=
\frac{M_{p,2}}{M_{p,1}}
\left(\frac{P_2}{P_1}\right)^{2/3}.
\label{eq:ratio_period_mass}
\end{equation}
Combining equations~(\ref{eq:eeff}) and (\ref{eq:ratio_period_mass}) gives
\begin{equation}
e_{\rm eff}
=
2\,\frac{M_{p,2}}{M_{p,1}}
\left(\frac{P_2}{P_1}\right)^{2/3}.
\label{eq:eeff_general}
\end{equation}
Equation~(\ref{eq:eeff_general}) shows that the apparent eccentricity of a single planet astrometric fit is not set solely by the inner planet's mass relative to the outer planet, but also by the geometric lever arm provided by the period (and therefore semi-major axis) ratio. For a strict 2:1 configuration with $P_2 = P_1/2$, we have
\begin{equation}
\label{eq:mratio_ecc}
\frac{M_{p,2}}{M_{p,1}}
= 2^{-1/3}\,e_{\rm eff} \simeq 0.79\,e_{\rm eff}.
\end{equation}

For instance, in this exact 2:1 case, an apparent eccentricity of $e_{\rm eff} = 0.30$ corresponds to $M_{p,2}/M_{p,1} \approx 0.24$, meaning that an inner planet with only one quarter of the mass of the outer planet can mimic the harmonic signature of a single planet with $e \approx 0.3$. Thus, the astrometric resoeccentric degeneracy maps cleanly into a mass-period relation, providing a physically intuitive explanation of how an apparently eccentric orbit can arise from two circular, phase-aligned planets in resonance. As eccentricity will always be less than unity, in order for two planets to mimic a single eccentric planet their mass ratio must be less than $\sim$0.79 with the inner planet having the smaller mass of the two planets.

\subsection{Mutual inclination breaks the degeneracy}
\label{sec:minc_degen}
If the two planets have different orbital orientations in the sky plane, 
$(\Omega_1,i_1,\omega_1)\neq(\Omega_2,i_2,\omega_2)$, they generate 
independent sets of Thiele-Innes constants 
$(A'_1,B'_1,F'_1,G'_1)$ and $(A'_2,B'_2,F'_2,G'_2)$.  
The sky-plane motion then takes the form
\begin{eqnarray}
\Delta\alpha^\ast &=& \frac{B'_1x_1(t) + G'_1y_1(t) + B'_2x_2(t) + G'_2y_2(t)}{d}, \\
\Delta\delta      &=& \frac{A'_1x_1(t) + F'_1y_1(t) + A'_2x_2(t) + F'_2y_2(t)}{d},
\end{eqnarray}
which is a sum of two ellipses with, in general, different orientations.  
A single Keplerian orbit possesses only one set of $(A',B',F',G')$, and therefore enforces fixed relationships among the harmonic coefficients in $\Delta\alpha^\ast$ and $\Delta\delta(t)$.  
Once the two planetary contributions have different projection geometries, those relationships no longer hold, and the motion cannot be reproduced by any single orbit.

For pairs of \emph{circular}-orbiting planets in a 2:1 configuration with $e_1=e_2=0$, the sky-plane coordinates depend on the mean anomaly and $\omega$ only through the mean longitude $\lambda = M + \omega$. A shift in $\omega$ can therefore be absorbed into a compensating shift in $M$, and differences in $\omega$ (or in $M$) do not create different Thiele-Innes constants for circular orbits.  Such phase shifts merely change the timing of the sinusoidal motion and do not alter its projected orientation.

Consequently, for circular-orbiting planets the degeneracy is broken \emph{only} by relative differences between the two planets' inclinations or longitudes of ascending node. Mutual inclination or nodal offsets produce distinct Thiele-Innes constants, therefore resolving the resoeccentric degeneracy. In contrast, perfectly coplanar circular planets with the same $(i,\Omega)$ remain degenerate with a single eccentric orbit at first order, regardless of differences in $\omega$ or in orbital phase.

The precise mutual inclination at which this degeneracy breaking will occur will depend on a number of factors -- including the masses and periods of the planets in the system, how far from a perfect 2:1 configuration are the two planets, the mass of the star, the distance to the star, and the location of the star in the sky (which affects the cadence and number of \textit{Gaia} observations) as well as modeling and sampling choices. As such, we leave a detailed analysis of this to future work.

\section{Simulations}\label{sec:simulations}

\begin{figure*}
    \centering
    \includegraphics[width=\textwidth]{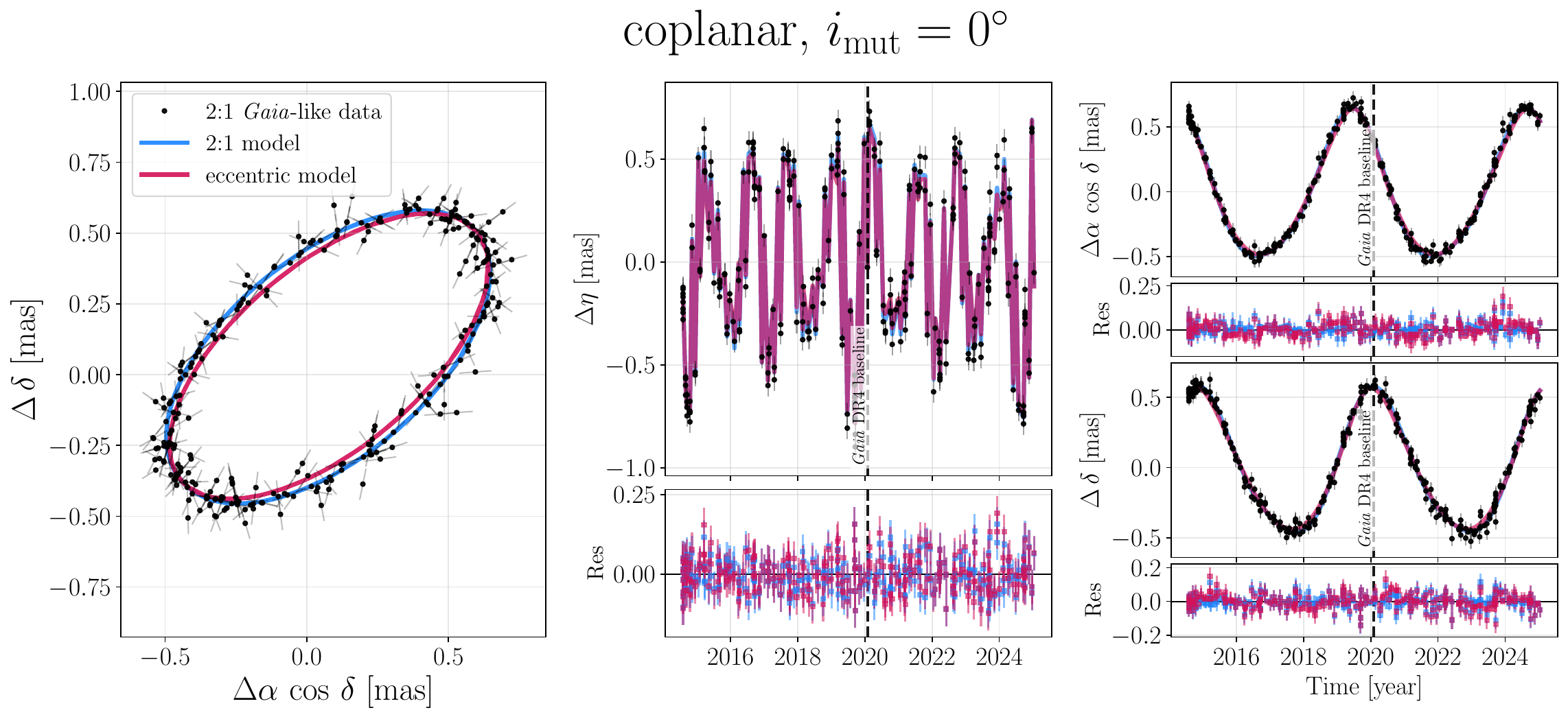}
    \caption{\textit{Gaia}-like simulated orbits (vertical line shows \textit{Gaia} DR4 baseline), coplanar system ($T_\mathrm{span}=5.5\,\mathrm{yr}$, $i_\mathrm{mut}=0^\circ$): 2:1 system with $P_1=5.2\,\mathrm{yr}$, $M_{p,1}=12\,M_\mathrm{Jup}$, $e_1=0.0$, $P_2=2.6\,\mathrm{yr}$, $M_{p,2}=2.84\,M_\mathrm{Jup}$, $e_2=0.0$, compared to a single planet model with $P=5.2\,\mathrm{yr}$, $e=0.3$, $M_p=12\,M_\mathrm{Jup}$. Star assumed to be solar-mass at 50\,pc. Here, $\Delta\alpha\cos\delta = \Delta\alpha^\ast$ is the RA and $\Delta\delta$ is the declination of the stellar reflex motion projected onto the sky plane of a star; $\Delta\eta$ is the \textit{Gaia} along–scan (AL) coordinate; Res are the residuals between the simulated \textit{Gaia}-like data with the eccentric model and the 2:1 model, respectively.}
    \label{fig:coplanar}
\end{figure*}

We demonstrate how \textit{Gaia} astrometric observations may have difficulties in differentiating between two-planet systems in a 2:1 mean-motion commensurability and a single eccentric planet. Using the \texttt{jaxoplanet} \citep{jaxoplanet} framework, we create a Keplerian orbit for a pair of $e=0$ giant planets with masses $M_{p,1}=12\,M_{\mathrm{Jup}}$ and $M_{p,2}=(0.79)\,(e_{\rm eff}=0.3)\,({M_{p,1}}) = 2.84\,M_{\mathrm{Jup}}$ and with orbital periods $P_1$ and $P_2=P_1/2$. Following the same procedure, we additionally create a Keplerian orbit for a single eccentric planet of mass $M_{p,1}=12\,M_{\mathrm{Jup}}$, period $P_1$, and eccentricity $e_\mathrm{eff} =  0.3$. We then duplicate this effort for a 2:1 system with a range of mutual inclinations set to $\{0, 15, 30, 45\}^\circ$ to demonstrate the resoeccentric degeneracy breaking capabilities of mutually inclined orbits.

We then simulate \textit{Gaia-like} data, using the full \textit{Gaia} scanning law via \texttt{gaiascanlaw}\footnote{\url{https://github.com/zpenoyre/gaiascanlaw}}\citep{gaiascanlaw} and adopting an along scan precision ($\sigma_\textrm{fov}$) of 54 $\mu$as -- a reasonable assumption for stars with G-band magnitude $<$ 14 \citep{Holl2023}. We chose an arbitrary stellar location (RA = $\alpha=20^\circ$, DEC = $\delta=-45^\circ$), set the stellar mass to a solar mass, and set the distance to 50 $pc$. 

\textit{Gaia} does not measure instantaneous RA and Dec, but rather the along–scan (AL) coordinate $\Delta\eta$ at each field-of-view transit. For every simulated epoch $t_k$ with scan angle $\psi_k$ (computed from the full \textit{Gaia} scanning law using \texttt{gaiascanlaw} \citep{gaiascanlaw}), we project the sky-plane reflex motion into the AL direction via
\[
\Delta\eta_k^{\rm true}
    = \Delta\alpha^\ast_k \cos\psi_k
    + \Delta\delta_k \sin\psi_k .
\]
We then add Gaussian noise with standard deviation $\sigma_{\rm fov}$ to obtain
\[
\Delta\eta_k^{\rm obs}
    = \Delta\eta_k^{\rm true}
    + \mathcal{N}(0,\sigma_{\rm fov}).
\]

The corresponding noisy sky-plane measurements follow by projecting the AL noise realization $\epsilon_k=\Delta\eta_k^{\rm obs}-\Delta\eta_k^{\rm true}$ back onto RA$^\ast$ and Dec:
\[
\Delta\alpha^\ast_{\rm obs,k}
  = \Delta\alpha^\ast_{\rm true,k} + \epsilon_k\cos\psi_k,
\]

\[
\Delta\delta_{\rm obs,k}
  = \Delta\delta_{\rm true,k} + \epsilon_k\sin\psi_k.
\]

The marginalized RA and Dec uncertainties are therefore
\[
\sigma_{\alpha^\ast,k} = \sigma_{\rm fov}\,|\cos\psi_k|,
\qquad
\sigma_{\delta,k}      = \sigma_{\rm fov}\,|\sin\psi_k|.
\]

\begin{figure*}
    \centering
    \includegraphics[width=\textwidth]{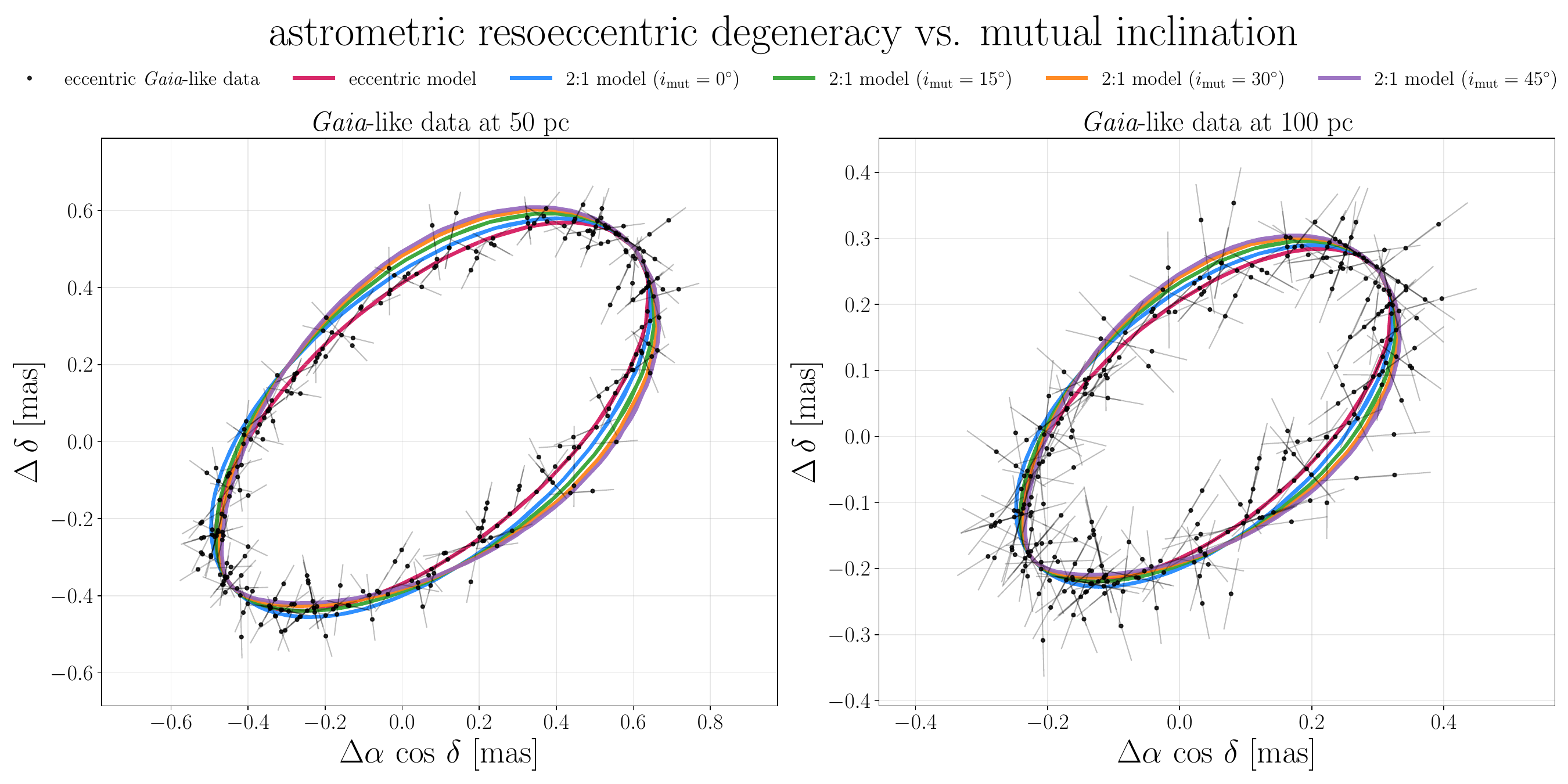}
    \caption{\textit{Gaia}-like simulated orbits showing that at some mutual inclination the astrometric resoeccentric degeneracy breaks. We simulate the same systems as in Figure~\ref{fig:coplanar}, but with mutual inclinations of $i_\mathrm{mut}=\{0, 15, 30, 45\}^\circ$ for a 2:1 system with $P_1=5.2\,\mathrm{yr}$, $M_{p,1}=12\,M_\mathrm{Jup}$, $e_1=0.0$,  $P_2=2.6\,\mathrm{yr}$, $M_{p,2}=2.84\,M_\mathrm{Jup}$, $e_2=0.0$,  compared to a single planet model with $P=5.2\,\mathrm{yr}$, $e=0.3$, $M_p=12\,M_\mathrm{Jup}$. Star assumed to be solar-mass at 50 $pc$ [left] or 100 $pc$ [right]. Here, $\Delta\alpha\cos\delta = \Delta\alpha^\ast$ is the RA and $\Delta\delta$ is the declination of the stellar reflex motion projected onto the sky plane of a star.}
    \label{fig:inclined}
\end{figure*}

With this approach, we can generate synthetic sky–plane (RA$^\ast$, Dec) reflex motion as well as along-scan ($\Delta\eta$) measurements for \textit{Gaia} DR4 (5.5\,yrs) and \textit{Gaia} DR5 (10.5\,yrs). The resulting sky-plane orbits, RA/Dec time series for these simulations, and along-scan
astrometric measurements ($\Delta\eta)$ are shown in Figure~\ref{fig:coplanar}. In Figure~\ref{fig:inclined}, we reproduce this effort showing only the sky–plane (RA$^\ast$, Dec) for a range of mutual inclinations ($i_\mathrm{mut}=\{0, 15, 30, 45\}^\circ$) for the 2:1 system and the host star at 50 $pc$ and 100 $pc$. These demonstrate that [1] it is possible for a single eccentric planet to mimic a coplanar, circular 2:1 resonant planetary system in both \textit{Gaia} DR4 and DR5 and [2] in certain orbital \& mass configurations, mutual inclination will break this degeneracy -- allowing observers to statistically distinguish between these two model hypotheses. We emphasize that these figures are only for a single planetary orbital-mass configuration and should not be taken as representative of the full population of \textit{Gaia} DR4 and DR5 planetary systems -- but rather a proof-of-concept for \textit{Gaia}-like data. Future work should investigate for which orbital configurations precisely will this degeneracy breaking occur in \textit{Gaia} DR4 and DR5.

\section{Implications}\label{sec:implications}
The astrometric resoeccentric degeneracy is a mathematical degeneracy that is perfect to first order in eccentricity and under the assumption of ``perfect data.'' In this section, we discuss some of the consequences of this degeneracy in the context of the imminent release of \textit{Gaia} DR4, as well as the dynamical histories implied by planets in 2:1 MMR configurations. While our discussions regarding the likely architectures of these 2:1 MMR systems rest upon an extensive legacy of past work, we briefly opine that \textit{Gaia} DR4 and DR5 will significantly improve our population-level understanding of this range of exoplanetary orbital-mass parameter space: just as other pioneering discoveries \citep[e.g.,][]{mayor1995eg} inspired revisions of theoretical outcomes of planet formation, the astrometric exoplanet revolution may similarly reshape current expectations. In this context, there is value in remaining open to a broader range of possible planetary architectures, recognizing that those emphasized in the literature largely reflect current observational capabilities.

\subsection{Exoplanet Astrometry in the \textit{Gaia} Era}
\textit{Gaia} is projected to discover $\sim$1,900 long-period planets with the release of DR4 and $\sim$38,000 with DR5  \citep[at $<$20\% mass and period precision;][]{lammers2025exoplanet}, predominantly at orbital separations around $2-5\,\mathrm{AU}$. Many more partial orbits and systems that cannot be as precisely constrained will also be observed \citep[e.g.,][]{Wallace2025} at even further orbital separations. While some giant planets on multi-AU orbits are currently known, this dataset will represent the first population-level probe of planetary systems at several astronomical units, providing a unique opportunity for analysis of demographics and occurrence rates of these long-period planets.

Our work highlights the meticulous care that must be taken in interpreting these demographics. We predict that if single planet solutions are assumed, an overabundance of eccentric planets may be reported. As we shall elaborate upon in \S \ref{sec:minc}, if these eccentric planets are incorrectly taken at face value, our understanding of their dynamical histories shall be led astray. If this degeneracy is not resolved by mutual inclination as discussed in \S \ref{sec:minc_degen}, we encourage community efforts to pursue follow-up campaigns and continue to develop rigorous sampling techniques to address it, as outlined in \S \ref{sec:break_degen}. 

We speculate that if an unusual overabundance of planets is indeed detected at a given eccentricity, it may be possible to draw conclusions regarding the occurrence of planets of the corresponding mass ratio (see Equation~\ref{eq:mratio_ecc}). We defer detailed analysis of this to future work.


\subsection{Long-Period Resonant Planets May Be Common}
\label{sec:chains}
In multi-planet systems, interactions with the protoplanetary disk generally induce convergent inward migration \citep{lin1986proto}, making compact resonant chains the expected outcome of planetary formation \citep[e.g.][]{snellgrove2001disc,LeePeale2002}. Indeed, the outer Solar System is believed to have originated in such a resonant configuration \citep{tsiganis2005, gomes2005, morbi2005}. While \cite{dai2024resonance} showed resonance is indeed common in young systems, this trend is not observed in more mature systems. The generally accepted picture is that resonant chains destabilized over the subsequent evolution of the system, a paradigm coined ``breaking the chains" \citep{izidoro2017chains}.

Many avenues for breaking close-in resonant chains have been proposed \citep[e.g.][]{lithwick2012resonant, batygin2017disruption, choksi2020neptune, hansen2024widespread, liu2025hydro}. However, many of these mechanisms are likely less efficient for disrupting long-period chains\footnote{We note that some mechanisms, such as stellar flybys \citep{charalambous2025breaking}, will be more efficient at disrupting long-period chains.} \citep[e.g.][]{batygin2013dissipative, nagpal2024breaking}. While the outer reaches of planetary systems are difficult to probe at present, the two best-characterized directly-imaged multi-planet systems HR 8799 \citep{wang2018hd} and PDS 70 \citep{bae2019pds} are consistent with resonant configurations. It hence may well be the case that long-period multi-planet resonant chains are a common system archetype. In fact, dynamical modeling of planet-planet scattering (see \S \ref{sec:minc}) for $\sim$Jupiter-mass systems that avoid close encounters suggests that this will be a common outcome ($\sim$50-80\% of systems) -- predicting that resonant chains should be frequent among massive planets in outer planetary systems \citep{Raymond2010}.

We note that migration of giant planets that have been captured into mean-motion resonance leads to the excitation of eccentricity, potentially to high values \citep[e.g.][]{LeePeale2002}. The non-zero but moderate eccentricities typically found in known systems of resonant giant planets \citep[e.g.][]{LeePeale2002, correia2009hd45364, rivera2010gj876} are consistent with this eccentricity growth mechanism being balanced against damping from planet-disk interactions \citep{papaloizou2000protoplanets, cresswell2007evolution}. For the high mass systems that will dominate the \textit{Gaia} population, simulations and analytic arguments show that the strength of damping due to the disk interaction may be substantially weaker, or even switch sign to become eccentricity growth \citep{papaloizou2001eccentricity,goldreich2003eccentricity,ogilvie2003eccentricity}. This could lead to substantially larger eccentricities for higher mass resonant pairs. 
A detailed analysis of the eccentricity limits in which this degeneracy holds is left to future work (see \S \ref{sec:break_degen} for more discussion). We note that these forthcoming \textit{Gaia} observations are expected to significantly improve population-level constraints on the eccentricity distribution of resonant giant planets, motivating consideration of a broad range of dynamical outcomes.

If many systems are discovered in \textit{Gaia} astrometry with very massive planets in 2:1 configurations, it would suggest that these planets likely formed in the disk and were then captured into resonance -- thus implying a very massive disk. This would be an interesting result in its own right and could potentially offer indirect constraints on planetary formation models.

\subsection{Mutually Inclined Systems}
\label{sec:minc}

The fact that mutual inclination can eliminate this degeneracy (see \S \ref{sec:minc_degen} and Figure~\ref{fig:inclined}) is perhaps fortuitous, given that the mutual inclination is also of paramount dynamical interest. Planets are expected to form in circular well-aligned orbits, as planet-disk interactions will damp out any eccentricity or inclination excursions \citep{cresswell2007evolution}. Hence, both eccentric planets and mutually inclined planets \citep[e.g.][]{xuan2020mutual,an2025significant,bgagliuffi2025jwst} may be signals of dynamically hot histories. However, the inferred evolutionary pathways differ for an isolated eccentric planet versus a system containing two misaligned planets. Below, we summarize the dynamical histories predicted by several mechanisms considered especially influential in sculpting exoplanetary architectures.

\begin{itemize}
    \item Planet-planet scattering \citep{chatterjee_2008,juric2008dynamical,nagasawa2008scattering,ford2008,petrovich2014scattering,anderson2020insitu,marzari2025planet,lu2025hatp11} involves violent gravitational interactions induced by close encounters between pairs of planets, and can generate both significant eccentricity and inclination. \cite{raymond2008mmrs} showed that while mean motion resonances are not the typical outcome of planet-planet scattering, this outcome does occur with appreciable frequency. They state

    \begin{quote}
        ``Thus, scattering is likely to be responsible for systems past 1 AU with 2:1 or 3:2 resonant planets and a more massive outer planet."
    \end{quote}
    
    Planet-planet scattering events involving only two planets tend to eject one of them, and rarely generate extremely high mutual inclinations in the cases where both planets survive \citep{ford2001dynamical, nagasawa2011orbital}. Hence, a mutually inclined system likely originated with three or more planets while a single eccentric planet could be generated with only two planets, if scattering is responsible.
    \item Secular interactions between planets can result in both excited eccentricities and inclinations \citep{MurrayDermott1999}, and these effects are greatly enhanced near mean motion resonances \citep[e.g.][]{malhotra1998resonances}. If instead a single eccentric planet must be explained via secular interactions, the dynamical interpretation changes -- the perturbing planet must be distant enough as to be undetectable. For instance, the system may be hierarchical and an unseen distant body may  induce ZLK oscillations \citep{vonzeipel1910zlk,lidov1962evolution,kozai1962secular,naoz2016eccentric}.
    \item Primordial misalignment. While eccentricity is very difficult to explain without invoking some sort of dynamical mechanism \citep[e.g.][]{duffell2015eccentric}, the same cannot be said for misalignments. Primordial misalignments can arise naturally through processes that tilt the protoplanetary disk itself \citep{lai2008wave, kuffmeier2021misaligned}, a scenario for which there is growing observational support \citep{BiddleBowler2025}. Intriguingly, however, we note that such observational efforts are only capable of probing the outer reaches of protoplanetary disks -- there is complementary evidence suggesting that inner disks are preferentially aligned \citep{radzom2024misalignment, radzom2025misalignment}. This supports the hypothesis that disks may often warp into two mutually misaligned sections \citep[e.g.][]{nixon2013tearing, nealon2018warp}, resulting in mutually inclined planets.
\end{itemize}
We note that dynamical mechanisms generally do not yield two perfectly circular orbits in a mean-motion resonance; some eccentricity is typically excited in the process. We emphasize that these are all first order approximations in eccentricity and thus will be increasingly less reliable as eccentricity grows. 

\subsection{Peas-in-a-Pod-like Architecture}
Our constraints on the mass ratios of distant giant planets are poor, both observationally and theoretically. A tempting assumption to make would be that of relatively equal-mass planets, a trend seen in the demographics of closer-in planets and coined ``peas-in-a-pod" \citep{millholland2017peas,weiss2018peas}.

If indeed the outer reaches of planetary systems largely exhibit the same intra-system mass uniformity as their close-in counterparts, the degeneracy we have pointed out will not be a concern. Inspection of \autoref{eq:mratio_ecc} reveals that should the two planets in question be of equal mass, a single planet masquerading as the pair would require an unphysical eccentricity of $e_\mathrm{eff}$ = 1.26 to mimic the astrometric signal. More generally, in systems where $M_{p,2} \gtrsim 0.79 M_{p,1}$ this degeneracy will not be a concern -- and in fact higher order terms in eccentricity will likely break the resoeccentric degeneracy at even smaller mass ratios.

\subsection{Hill-stability mass limits}
The preferred parameter space of \textit{Gaia}-detectable planets will be super-Jupiters on $2-5\,AU$ orbits \citep{lammers2025exoplanet}. At first glance, this narrow semimajor-axis range may seem too restrictive to host multiple massive planets on long-term stable orbits. In this section, we demonstrate that such configurations are in fact readily achievable.

The Hill stability boundary of two planets on initially coplanar circular orbits is well understood. \citet{Gladman1993} derived an analytic criterion for the orbital spacing $\Delta \equiv a_2 - a_1$ demanded for Hill stability for two planets of masses $M_{p,1}$ and $M_{p,2}$ orbiting a star of mass $M_\star$:

\begin{equation}
\label{eq:hillstab}
\frac{\Delta}{a_1}
> 
2.4\,\left(\frac{M_{p,1}+M_{p,2}}{M_\star}\right)^{1/3}.
\label{eq:gladman}
\end{equation}
We derive the maximum total planetary mass compatible with Hill stability for a pair of planets locked in a 2:1 period commensurability. From Kepler's third law \citep{kepler1609third}, $\Delta \approx 0.587a_1$ for the 2:1 resonance. We may substitute this into \autoref{eq:hillstab} and solve for the total planetary mass $M_{p,1} + M_{p,2}$:
\begin{equation}
M_{p,1}+M_{p,2}
\lesssim 0.0149 M_\star.
\end{equation}
For a solar-mass host star, two planets with a total mass of up to $\sim15.6 \,M_{\rm Jup}$, easily detectable by \textit{Gaia}, can be expected to be Hill stable around a solar mass star in a 2:1 period configuration -- thus further emphasizing the importance for observers to prepare for the possibility of this degeneracy. 

We note that \citet{Gladman1993} did not consider the stabilizing influence of mean motion resonances in their study. Numerous authors \citep[e.g.][]{wisdom1980resonance, marti2013gj876,obertas_2017,tamayo2017convergent,Tamayo20,lammers2024instability,lammers_winn_2024,lu_hip,tamayo2025unified} have demonstrated that exact mean motion resonances are associated with strong islands of stability. We present evidence in \S \ref{sec:chains} that capture into these stable islands via convergent migration is an expected outcome of planet formation. Additionally, it is worth noting that for giant planets, the formation of resonances is likely to occur while the protoplanetary disk is present. In this case, tidal damping adds an extra stabilizing factor not considered in the analysis by \citet{Gladman1993}. Altogether, this implies that even high-mass planets may also be stable on long timescales in 2:1 configurations.

\subsection{Other Ways to Break this Degeneracy}
\label{sec:break_degen}
The mathematical derivations in \S\ref{sec:math} are first order approximations for the single eccentric planetary case and assume circular and coplanar orbits for the 2:1 resonant case. There is thus some eccentricity and mutual inclination at which this approximation will no longer hold. This will depend on a number of factors, including: orbital parameters (most dominantly masses and periods) of the planet(s) in the system; observational limitations, which will also depend on the distance to, location of, and brightness of the host star; and modeling choices. Additionally, in these derivations, we have assumed Keplerian orbits. As such, we leave an in depth analysis of the eccentricities and mutual inclinations at which the degeneracy is likely to be broken for systems in \textit{Gaia} DR4 and DR5 to future work. In what follows, we discuss some possible strategies for mitigating and properly accounting for the resoeccentric degeneracy in \textit{Gaia} astrometry.

A transit observation could break this degeneracy, as an additional constraint on the eccentricity can be derived from the transit alone via the photo-eccentric effect \citep{Kipping2012, DawsonPhotoEccentric2012}. However, as transits are more sensitive to close-orbiting planets, we anticipate a minimal overlap between the population of \textit{Gaia} astrometric planets and systems with observed transits. We can estimate the number of \textit{Gaia} astrometric planets to transit by adopting the $\sim$1,900 planets in DR4 and $\sim$38,000 planets in DR5 with 20\% mass and period precision from \citet{lammers_winn_2024}. We use the planet transit probability formula assuming a circular orbit: $\mathcal{P}_{\text{tr}} = \frac{R_\star + R_p}{a}$ where $R_\star$ is the stellar radius, $R_p$ is the planetary radius, and $a$ is the semi-major axis. For a Jupiter radius planet orbiting a solar radius star at $2-5\,AU$, this equates to a transit probability of $\sim0.21\%-0.084\%$. For simplicity, we adopt the mean transit percentage ($\sim0.15\%)$ and predict that $\sim$3 \textit{Gaia} DR4 planets will transit and $\sim$57 \textit{Gaia} DR5 planets will transit. Any such systems will be of particular interest to better understand this degeneracy in \textit{Gaia} astrometry, and may require targeted follow-up to obtain these transit observations.

Significant effort has been dedicated to disentangling this resoeccentric degeneracy in radial velocity observations. As previously noted, \citet{AngladaEscude2010} originally identified that $\sim$35\% of single eccentric RV planets are statistically compatible with circular two-planet 2:1 configurations. \citet{Wittenmyer2012} presented an additional planet in this 2:1 configuration in two system previously identified as single eccentric planet systems. \citet{Wittenmyer2013} investigated 82 known (at the time) moderately eccentric single planet systems ($e > 0.3$) to test the hypothesis that some may actually be low-eccentricity two-planet systems -- identifying 9 particular systems of interest. A Bayesian model comparison of 60 moderately eccentric single planet systems with RV data identified 15 systems that favor 2:1 resonant near-circular orbits at 95\% confidence \citep{Boisvert2018}. \citet{Wittenmyer2019b} showed that RV-observed systems with two near-resonant planets on near-circular orbits most often masquerade as a single planet system with orbital eccentricity of approximately $e = 0.31 \pm 0.12$ in sparsely sampled data and that such systems were highly unlikely to imitate single planet systems with orbital eccentricities greater than $e \sim 0.5$. Presumably there will be a similar eccentricity at which this degeneracy will be most impactful and a maximum eccentricity at which this degeneracy will break in \textit{Gaia} astrometry. We leave an investigation into these eccentricity limits to future work.

Combining radial velocity and astrometric observations is a powerful way to determine true planetary masses and orbital inclinations. Radial velocities measure only the line-of-sight motion and therefore yield the minimum mass, $M_p \sin i$, while astrometry measures the sky-plane component of the stellar reflex motion. When used together, these complementary data sets break the longstanding $\sin i$ degeneracy of radial velocity analyses \citep[e.g.,][]{WrightHoward2009, Brandt2019, Winn2022, Yahalomi2023, Feng2024}. This RV-astrometry synergy has recently been demonstrated in practice via follow-up radial velocity monitoring to directly confirm and refine astrometric orbit solutions from \textit{Gaia} DR3. In particular, RV follow-up of the Gaia-4b and Gaia-5b candidates revealed a massive planet and a brown dwarf orbiting low-mass stars, validating the astrometric orbits and providing precise dynamical masses \citep{stefansson2025gaia}. We note that even the Gaia-4b and Gaia-5b signals -- with effective eccentricities $e_{\rm eff} \simeq 0.34$ and $e_{\rm eff} \simeq 0.64$, respectively, from the single planet fits in \citet{stefansson2025gaia} -- could in principle be mimicked by circular 2:1 planet pairs: for Gaia-4, a pair with periods $\{571, 286\}\,\mathrm{days}$ and masses $\{11.8, 3.2\}\,M_{\rm Jup}$; and for Gaia-5, a pair with periods $\{359, 179\}\,\mathrm{days}$ and masses $\{20.9, 10.6\}\,M_{\rm Jup}$. While a detailed investigation of the resoeccentric degeneracy for these two systems is beyond the scope of this work, such an analysis would help clarify whether, and under what conditions, RV follow-up of astrometrically identified systems can break this degeneracy. Future work should additionally investigate whether RV breaking of the resoeccentric degeneracy is feasible for the types of planets we expect to find in \textit{Gaia} DR4 and DR5 astrometry based on current RV capabilities and study the optimal orbital strategies for such observations.

This work predicts an additional degeneracy that may appear in the \textit{Gaia} epoch astrometric data. It is already well-documented that astrometric solutions become highly degenerate when the orbital period extends beyond the observational baseline \citep[e.g.,][]{Casertano2008}. This highlights the need for the development of proper astrometric inference techniques that are  designed to explore highly degenerate and multi-modal parameter spaces.  Modern orbit-fitting frameworks such as \texttt{The~Joker} \citep{thejoker}, \texttt{orbitize!} \citep{orbitize1, orbitize2}, OFTI \citep{ofti}, \texttt{Octofitter} \citep{octofitter}, and \texttt{orvara} \citep{orvara} have demonstrated the importance of using samplers that remain robust in the presence of strong covariances, long-lived degeneracies, and disconnected orbital solutions.  Techniques such as rejection sampling, parallel tempering, and nested sampling have been developed precisely for this purpose, providing reliable posterior exploration even when the likelihood surface contains multiple isolated modes or narrow curved manifolds. These approaches are essential for accurately capturing the full family of orbital solutions implied by astrometric and radial-velocity data, and we strongly encourage their further development and study in preparation for \textit{Gaia} DR4 and DR5. It will be important to understand in which regions of parameter space these different sampling methods optimally sample the complete solution space.

We also encourage future work to investigate the signal-to-noise ratio (SNR) required to differentiate between the 2:1 and eccentric configurations. This will likely be a function of the masses, periods, and other orbital parameters (such as eccentricity and mutual inclination of the 2:1 planets), but will be a powerful result in ensuring over what regions of parameter space are statistically differentiated solutions possible. See \citet{kippingRVsnr2024} for the analogous derivation of this for RV data, which was built on the inclusion of higher order eccentricity terms as derived in \citet{Lucy2005}.

Finally, it may be possible to detect this resoeccentric degeneracy at the population level by searching for non-isotropic solutions of $\omega$ (the argument of the periastron) of single eccentric planets fit to the \textit{Gaia} data. As the $\omega$ term would need to be ``fine-tuned'' in order to mimic the astometric signal of a 2:1 planetary configuration, the presence of unresolved resonant pairs would be expected to induce anisotropis in the inferred $\omega$ distribution. Such anisotropies could therefore serve as a population-level indicator of this degeneracy. Injection-recovery tests of 2:1 systems fit with single eccentric solutions could be used to validate this degeneracy-flagging approach prior to \textit{Gaia} DR4.

\section{Conclusion}
In this work, we derived the astrometric resoeccentric degeneracy: an extension of the 2:1-eccentric degeneracy familiar from RV studies. To first order in eccentricity, the sky-plane motion of a star induced by a single eccentric planet decomposes into a fundamental frequency and a first harmonic with an amplitude that scales with its eccentricity; a coplanar pair of circular planets in a 2:1 period ratio induces the same harmonic structure. We provide a simple mass-eccentricity mapping that links the two descriptions, defining exactly when a two-planet system can masquerade as a single eccentric one and vice-versa.

Using \textit{Gaia}-like DR4 and DR5 simulations, we showed that the reflex motion induced by a mildly eccentric single planet could mimic the reflex motion of a circular 2:1 planetary pair. We also demonstrate that mutual inclination can break this degeneracy by introducing distinct projection geometries for the two planets. 

If single planet fits are assumed by default, circular resonant pairs may be mistaken for single eccentric planets -- biasing inferred eccentricity distributions, dynamical interpretations, and occurrence rates. This underscores the need for orbit-fitting approaches that can handle multi-modal astrometric likelihoods, including this resoeccentric degeneracy. Samplers capable of robustly exploring degenerate orbital solutions are essential to avoid collapsing 2:1 configurations into single eccentric solutions. A combined astrometry and RV approach appears particularly promising. Systems like Gaia-4b and Gaia-5b -- where RV follow-up already exists -- offer ideal real-world test beds for assessing when and how RV observations can break the 2:1-eccentric astrometric degeneracy.

Several natural extensions follow from this work. It will be important to map out the regions of parameter space in which mutual inclination or follow-up measurements reliably distinguish the two orbital architectures, and to quantify the signal-to-noise thresholds neede for \textit{Gaia} to do the same. Population-level indicators, such as anisotropies in the fitted argument of periastron distribution, may provide additional diagnostic power. Injection-recovery experiments that include resonant multi-planet architectures -- in addition to single planet systems -- will be key for unbiased occurrence-rate measurements.

Taken together, these efforts will help ensure that \textit{Gaia}'s epoch astrometry yields not just the detection of many long-period giant planets but an accurate understanding of their occurrence rates, architectures, and the dynamical processes that shaped them.

\begin{acknowledgments}
We thank the anonymous reviewer who's feedback improved this manuscript. We are very grateful to David Kipping for comments and feedback on an earlier version of this manuscript. We thank the CCA Astronomical Data Group and the Rice Research Group at Yale University for helpful and illuminating conversations. D.A.Y. is supported by a Juan Carlos Torres Postdoctoral Fellowship at the Massachusetts Institute of Technology. D.A.Y. and T.L. are supported by Flatiron Research Fellowships at the Flatiron Institute, a division of the Simons Foundation. M.R. acknowledges support from Heising-Simons grants \#2021-2802 and \#2023-4478, as well as National Geographic grant EC-115062R-24 and NASA Exoplanets Research Program NNH23ZDA001N-XRP (grant No. 80NSSC24K0153). ChatGPT was utilized to improve wording at the sentence level and assist with coding inquires – last accessed
in 2026 January.
\end{acknowledgments}

\software{This work made use of the following software packages: \texttt{matplotlib} \citep{Hunter:2007}, \texttt{numpy} \citep{numpy}, \texttt{python} \citep{python}, \texttt{JAX} \citep{jax2018github}, \texttt{jaxoplanet} \citep{jaxoplanet}, and \texttt{gaiascanlaw} \citep{gaiascanlaw} Software citation information aggregated using \texttt{\href{https://www.tomwagg.com/software-citation-station/}{The Software Citation Station}} \citep{software-citation-station-paper,software-citation-station-zenodo}.}


\bibliography{main}{}
\bibliographystyle{aasjournalv7}


\end{CJK*}
\end{document}